\documentclass[pra]{revtex4}
 \usepackage{amssymb} \usepackage{graphicx}
\usepackage{chemarrow}

\begin{document}
 \title{Harmonic spinors from twistors and potential forms}

\author{\"Umit Ertem}
 \email{umit.ertem@tedu.edu.tr, umitertemm@gmail.com}
\address{TED University, Ziya G\"{o}kalp Caddesi No:48, 06420, Kolej \c{C}ankaya, Ankara, Turkey\\}

\begin{abstract}

Symmetry operators of twistor spinors and harmonic spinors can be constructed from conformal Killing-Yano forms. Transformation operators relating twistors to harmonic spinors are found in terms of potential forms. These constructions are generalized to gauged twistor spinors and gauged harmonic spinors. The operators that transform gauged twistor spinors to gauged harmonic spinors are found. Symmetry operators of gauged harmonic spinors in terms of conformal Killing-Yano forms are obtained. Algebraic conditions to obtain solutions of the Seiberg-Witten equations are discussed.

\end{abstract}

\maketitle

\section{Introduction}

In a spin manifold $M$, two basic first-order differential operators can be defined on the spinor bundle which are Dirac operator and twistor operator. They are defined as projections on the tensor product of the tangent bundle and spinor bundle and they can be written in terms of the Clifford multiplication and the Levi-Civita connection \cite{Baum Friedrich Grunewald Kath,Bourguignon et al}. The spinor fields that are in the kernels of these operators have importance in mathematical physics. The set of spinor fields which are in the kernel of the Dirac operator corresponds to harmonic spinors and they are solutions of the massless Dirac equation. Twistor spinors are in the kernel of the twistor operator and they satisfy the twistor equation which appear in the supersymmetric field theories as the equation satisfied by the supersymmetry generators \cite{de Medeiros,Cassani Klare Martelli Tomasiello Zaffaroni}. Moreover, the common property of the massless Dirac equation and the twistor equation is that they have conformal covariance. So, the conformal symmetries of the ambient manifold is related to harmonic spinors and twistor spinors. The spinor bilinears of twistor spinors correspond to conformal Killing-Yano (CKY) forms which are antisymmetric generalizations of conformal Killing vector fields to higher degree forms \cite{Acik Ertem1}. CKY forms are also used in the construction of symmetry operators of harmonic and twistor spinors \cite{Benn Charlton,Ertem1,Ertem2}. Symmetry operators of an equation are defined as the operators that take any solution of the equation and give another solution. On the other hand, an operator that transform twistor spinors to harmonic spinors is proposed in \cite{Benn Kress} in terms of potential forms which are generalizations of the solutions of the generalized Laplace equation to higher degree forms.

Dirac and twistor operators can be generalized to manifolds with $Spin^c$ structure. A $Spin^c$ structure determines a bundle over the manifold whose sections correspond to $U(1)$-valued spinors. The Dirac operator constructed from the generalized connection induced on the $Spin^c$ bundle and includes the $U(1)$ gauge connection is called the gauged Dirac operator. The spinors which are in the kernel of the gauged Dirac operator are called gauged harmonic spinors. Similarly, gauged twistor operator can also be defined in the same way and its kernel consists of gauged twistor spinors. The massless gauged Dirac equation appears in the set of Seiberg-Witten equations which also correspond to the field equations of $N=2$ supersymmetric Yang-Mills theories \cite{Seiberg Witten}. Gauged harmonic spinors that satisfy an extra algebraic condition correspond to the solutions of the Seiberg-Witten equations and determine topological invariants in four-dimensional manifolds. Gauged twistor spinors also play important roles in supersymmetric field theories, since they generate the preserved supersymmetries of supersymmetric and superconformal field theories in curved backgrounds \cite{Festuccia Seiberg,Klare Tomasiello Zaffaroni,Cassani Martelli}. Although the symmetry operators of gauged twistor spinors are obtained in constant curvature manifolds in \cite{Ertem3}, the similar construction for gauged harmonic spinors is still an open problem. Moreover, the general transformation operators which can determine relations between gauged twistor spinors and gauged harmonic spinors are also not known.

In this paper, we start by considering twistor spinors and harmonic spinors and construct the transformation operators relating them. We give the explicit proofs of the constructions in terms of generalized Laplace functions and also of potential forms in $n$-dimensions. In that way, we find the first-order, second-order and third-order transformation operators, which all reduce to first-order ones, from twistor spinors to harmonic spinors. Then, we continue to the investigation with gauged twistor spinors and gauged harmonic spinors. We prove that the transformation operators from gauged twistor spinors to gauged harmonic spinors can be constructed in terms of generalized gauged Laplace functions and gauged potential forms in constant curvature manifolds. We also show that the symmetry operators of gauged harmonic spinors can be written in terms of CKY forms in constant curvature manifolds. As a result, we construct symmetry operators of gauged twistor spinors and gauged harmonic spinors and the first-order, second-order and third-order transformation operators, which all reduce to first-order ones, relating them in constant curvature manifolds. After that, we find the explicit algebraic conditions to obtain Seiberg-Witten solutions from gauged harmonic spinors in terms of symmetry and transformation operators which are written by using CKY forms and gauged potential forms in four dimensions.

The paper is organized as follows. In section II, we give the basic definitions and equalities which are used throughout the text. Section III contains the construction of the symmetry operators of twistor spinors and harmonic spinors. The proofs of the transformation operators in terms of generalized Laplace functions and potential forms are also given. In section IV, gauged twistor spinors and gauged harmonic spinors are considered and the proofs of the transformation operators from gauged twistor spinors to gauged harmonic spinors and the symmetry operators of gauged harmonic spinors are shown. Section V includes the algebraic conditions to obtain Seiberg-Witten solutions in four dimensions from the gauged harmonic spinors and transformation operators. Section VI concludes the paper.

\section{Preliminaries}

In this section, we start with the basic definitions and identities used in the proofs throughout the text \cite{Benn Tucker,Charlton,Acik Ertem Onder Vercin}. Let us consider an $n$-dimensional manifold $M$ with orthonormal frame basis $\{X_a\}$ and the dual co-frame basis $\{e^a\}$ with the property $e^a(X_b)=\delta^a_b$ where $\delta^a_b$ denotes the Kronecker delta symbol and $a,b=1,...,n$. The exterior bundle $\Lambda M$ of differential forms on $M$ gains a Clifford algebra structure with the following identity
\begin{equation}
e^a.e^b+e^b.e^a=2g^{ab}
\end{equation}
where $.$ denotes the Clifford product and $g^{ab}$ are the components of the metric tensor on $M$. With this algebra structure, the bundle of differential forms correspond to the Clifford bundle $Cl(M)$ on $M$ and the sections of this bundle are called Clifford forms. The Clifford product of a 1-form $x$ and a $p$-form $\alpha$ can be written in terms of the wedge product $\wedge$ and the interior derivative $i_X$ that corresponds to the contraction with respect to the vector field $X$ as follows
\begin{eqnarray}
x.\alpha&=&x\wedge\alpha+i_{\widetilde{x}}\alpha\\
\alpha.x&=&x\wedge\eta\alpha-i_{\widetilde{x}}\eta\alpha
\end{eqnarray}
where $\widetilde{x}$ is the metric dual of $x$ and $\eta$ is the main automorphism of the exterior algebra that acts on a $p$-form $\alpha$ as $\eta\alpha=(-1)^p\alpha$. In general, the Clifford forms are inhomogeneous differential forms and the Clifford bracket of two Clifford forms $\alpha$ and $\beta$ is written in terms of the wedge product as
\begin{eqnarray}
[\alpha, \beta]_{Cl}&=&\alpha.\beta-\beta.\alpha\nonumber\\
&=&\sum_{k=0}^{n}\frac{(-1)^{\lfloor{k/2\rfloor}}}{k!}\Bigg\{\left(\eta^k i_{X_{I(k)}}\alpha\right)\wedge i_{X^{I(k)}}\beta-\left(\eta^k i_{X_{I(k)}}\beta\right)\wedge i_{X^{I(k)}}\alpha\Bigg\}
\end{eqnarray}
where $I(k)=\{a_1, a_2, ... , a_k\}$ denotes a multi-index, $i_{X_{I(k)}}=i_{X_{a_1}}i_{X_{a_2}}...i_{X_{a_k}}$ and $\lfloor\rfloor$ denotes the floor function. Especially, if $\alpha$ is a homogeneous 2-form, then the Clifford bracket simplifies to
\begin{equation}
[\alpha,\beta]_{Cl}=-2i_{X^a}\alpha\wedge i_{X_a}\beta.
\end{equation}
Here and throughout the text, we use the Einstein summation convention. Moreover, the Clifford product has the following property for a $p$-form $\alpha$
\begin{equation}
e^a.\alpha.e_a=(-1)^p(n-2p)\alpha.
\end{equation}

The Levi-Civita connection $\nabla$ on the Clifford bundle determines a covariant derivative $\nabla_X$ with respect to the vector field $X$ which is compatible with the Clifford product
\begin{equation}
\nabla_X(\alpha.\beta)=\nabla_X\alpha.\beta+\alpha.\nabla_X\beta
\end{equation}
for $\alpha$ and $\beta$ are arbitrary Clifford forms. The exterior derivative operator $d$ and the co-derivative operator $\delta$ can be written in terms of covariant derivative as
\begin{equation}
d=e^a\wedge\nabla_{X_a}\quad\quad,\quad\quad\delta=-i_{X^a}\nabla_{X_a}
\end{equation}
for zero torsion. By using them, the Hodge-de Rham operator $\displaystyle{\not}d$ is defined on Clifford forms in the following form
\begin{equation}
\displaystyle{\not}d=e^a.\nabla_{X_a}=d-\delta.
\end{equation}

From the connection on the Clifford bundle, the curvature operator is written as
\begin{equation}
R(X,Y)=[\nabla_X,\nabla_Y]-\nabla_{[X,Y]}
\end{equation}
for vector fields $X$ and $Y$. The action of the curvature operator on an arbitrary Clifford form $\alpha$ can be written in terms of the curvature 2-forms $R_{ab}$
\begin{eqnarray}
R(X_a,X_b)\alpha&=&\frac{1}{2}[R_{ab}, \alpha]_{Cl}\nonumber\\
&=&-i_{X^c}R_{ab}\wedge i_{X^c}\alpha.
\end{eqnarray}
The other curvature characteristics defined from the curvature 2-forms are Ricci 1-forms $P_a=i_{X^b}R_{ba}$ and the curvature scalar ${\cal{R}}=i_{X^a}P_a$. They satisfy the following identities for zero torsion
\begin{equation}
R_{ab}\wedge e^b=0\quad\quad,\quad\quad P_a\wedge e^a=0.
\end{equation}
The curvature endomorphism $I(R)$ is defined from the curvature operator as
\begin{equation}
I(R)=e^a\wedge i_{X^b}R(X_b,X_a)
\end{equation}
and its action on a Clifford form $\alpha$ is
\begin{eqnarray}
I(R)\alpha&=&\frac{1}{4}R_{ab}.\alpha.e^{ab}+\frac{1}{4}{\cal{R}}\alpha\\
&=&P_a\wedge i_{X^a}\alpha-R_{ab}\wedge i_{X^b}i_{X^a}\alpha\nonumber.
\end{eqnarray}
Moreover, the square of the Hodge-de Rham operator defined in (9) can be written in terms of the curvature endomorphism as follows
\begin{equation}
\displaystyle{\not}d^2=\nabla^2-I(R)
\end{equation}
where $\nabla^2$ is the trace of the Hessian
\begin{equation}
\nabla^2=\nabla_{X^a}\nabla_{X_a}-\nabla_{\nabla_{X^a}X_a}
\end{equation}
and $\displaystyle{\not}d^2$ corresponds to the Laplace operator $\Delta$
\begin{equation}
\Delta=\displaystyle{\not}d^2=-d\delta-\delta d.
\end{equation}

In a manifold with a spin structure, a spinor bundle $S(M)$ on $M$ can be induced from the Clifford bundle and the sections of the spinor bundle are called spinor fields. Since the spinors correspond to the elements of the minimal left ideal of the Clifford algebra, the Clifford forms act on spinor fields by the Clifford multiplication from the left. The connection defined on the Clifford bundle in (7) induces a connection on the spinor bundle which we also denote as $\nabla$ and is compatible with the Clifford product of two arbitrary spinor fields $\phi$ and $\psi$
\begin{equation}
\nabla_X(\phi.\psi)=\nabla_X\phi.\psi+\phi.\nabla_X\psi.
\end{equation}
From the spinor covariant derivative, the Dirac operator $\displaystyle{\not}D$ acting on spinors is defined as follows
\begin{equation}
\displaystyle{\not}D=e^a.\nabla_{X_a}.
\end{equation}

The curvature operator on spinor fields is also defined as in (10) and its action on a spinor field $\psi$ is given by
\begin{equation}
R(X_a, X_b)\psi=\frac{1}{2}R_{ab}.\psi.
\end{equation}
The square of the Dirac operator acting on an arbitrary spinor can be written as
\begin{equation}
\displaystyle{\not}D^2=\nabla^2-\frac{1}{4}{\cal{R}}.
\end{equation}
The spinor fields that are in the kernel of the Dirac operator are called harmonic spinors and satisfy the following massless Dirac equation
\begin{equation}
\displaystyle{\not}D\psi=0
\end{equation}
and similarly, the eigenspinors of the Dirac operator with non-zero eigenvalue correspond to the solutions of the massive Dirac equation
\begin{equation}
\displaystyle{\not}D\psi=m\psi
\end{equation}
where $m$ is a real number. 

If a manifold $M$ has a $Spin^c$ structure, then one can define the $Spin^c$ bundle $S^c(M)$ whose sections correspond to $U(1)$-valued spinor fields. The gauged connection on $S^c(M)$ denoted by $\widehat{\nabla}$ is defined from the Levi-Civita connection $\nabla$ and the gauge connection 1-form $A$ as
\begin{equation}
\widehat{\nabla}_X=\nabla_X+i_XA
\end{equation}
for an arbitrary vector field $X$. One can choose normal coordinates $\{X_a\}$ that the gauged connection satisfies the following identities
\begin{equation}
\widehat{\nabla}_{X_a}X_b=0=[X_a,X_b].
\end{equation}
Similar to the operators $d$ and $\delta$ defined in (8), the gauged exterior derivative and gauged co-derivative operators are defined as
\begin{equation}
\widehat{d}=e^a\wedge\widehat{\nabla}_{X_a}\quad\quad,\quad\quad\widehat{\delta}=-i_{X^a}\widehat{\nabla}_{X_a}
\end{equation}
and they can be written in terms of $d$ and $\delta$ in the following form
\begin{eqnarray}
\widehat{d}&=&d+A\wedge\\
\widehat{\delta}&=&\delta-i_{\widetilde{A}}
\end{eqnarray}
where $\widetilde{A}$ is the metric dual of the 1-form $A$. Similarly, gauged Hodge-de Rham operator acting on Clifford forms is written as
\begin{equation}
\widehat{\displaystyle{\not}d}=e^a.\widehat{\nabla}_{X_a}=\widehat{d}-\widehat{\delta}\nonumber=\displaystyle{\not}d+A.
\end{equation}
However, unlike the case of $d$ and $\delta$ that satisfy $d^2=\delta^2=0$, the squares of gauged exterior and gauged co-derivatives are written in terms of the gauge curvature 2-form $F=dA$ as follows
\begin{eqnarray}
\widehat{d}^2&=&F\wedge\\
\widehat{\delta}^2&=&-(i_{X^a}i_{X^b}F)i_{X_a}i_{X_b}
\end{eqnarray}
and the gauged Laplace operator is defined as
\begin{equation}
\widehat{\Delta}=\widehat{\displaystyle{\not}d}^2=(\widehat{d}-\widehat{\delta})^2.
\end{equation}

Gauged curvature operator is defined from the gauged connection as
\begin{equation}
\widehat{R}(X,Y)=[\widehat{\nabla}_X,\widehat{\nabla}_Y]-\widehat{\nabla}_{[X,Y]}
\end{equation}
and it can be written in terms of the curvature operator in (10) and the gauge curvature $F$ as follows
\begin{equation}
\widehat{R}(X_a,X_b)=R(X_a,X_b)-i_{X_a}i_{X_b}F.
\end{equation}

Gauged Dirac operator $\widehat{\displaystyle{\not}D}$ acting on $Spin^c$ spinors is written in terms of the ordinary Dirac operator as
\begin{equation}
\widehat{\displaystyle{\not}D}=e^a.\widehat{\nabla}_{X_a}=\displaystyle{\not}D+A
\end{equation}
and the square of the gauged Dirac operator can be written in the following form
\begin{equation}
\widehat{\displaystyle{\not}D}^2=\widehat{\nabla}^2-\frac{1}{4}{\cal{R}}+F.
\end{equation}
Gauged harmonic spinors correspond to the spinor fields which are in the kernel of the gauged Dirac operator and satisfy the massless gauged Dirac equation
\begin{equation}
\widehat{\displaystyle{\not}D}\psi=0
\end{equation}
and the eigenspinors of the gauged Dirac operator are solutions of the massive gauged Dirac equation
\begin{equation}
\widehat{\displaystyle{\not}D}\psi=m\psi
\end{equation}
with $m$ a real number.

\section{Twistors to harmonic spinors}

In an $n$-dimensional spin manifold $M$, one can define two different first-order differential operators on spinor fields. The first one is the Dirac operator defined in (19) and the second one is the Penrose operator written as follows
\begin{equation}
{\cal{P}}_X:=\nabla_X-\frac{1}{n}\widetilde{X}.\displaystyle{\not}D
\end{equation}
with respect to a vector field $X$ and its metric dual $\widetilde{X}$. The spinor fields that are in the kernel of the Penrose operator are called twistor spinors and they satisfy the following twistor equation
\begin{equation}
\nabla_X\psi=\frac{1}{n}\widetilde{X}.\displaystyle{\not}D\psi
\end{equation}
for a spinor $\psi$. By applying the covariant derivative operation to the twistor equation, one obtains the integrability conditions of the twistor equation and the existence of twistor spinors restrict the curvature characteristics of the manifold. The integrability conditions of the twistor equation are obtained as follows \cite{Baum Friedrich Grunewald Kath}
\begin{eqnarray}
\displaystyle{\not}D^2\psi&=&-\frac{n}{4(n-1)}{\cal{R}}\psi\\
\nabla_{X_a}\displaystyle{\not}D\psi&=&\frac{n}{2}K_a.\psi\\
C_{ab}.\psi&=&0
\end{eqnarray}
where the 1-form $K_a$ is
\begin{equation}
K_a=\frac{1}{n-2}\left(\frac{\cal{R}}{2(n-1)}e_a-P_a\right)
\end{equation}
and $C_{ab}$ is the conformal 2-forms defined as
\begin{equation}
C_{ab}=R_{ab}-\frac{1}{n-2}\left(P_a\wedge e_b-P_b\wedge e_a\right)+\frac{1}{(n-1)(n-2)}{\cal{R}}e_{ab}.
\end{equation}
Here, $e_{ab}$ denotes $e_a\wedge e_b$. Eq. (43) implies that twistor spinors can exist on conformally-flat manifolds.

Solutions of the twistor equation from a known solution can be obtained by using the symmetry operators. It is known that the Lie derivative ${\cal{L}}_K$ with respect to a conformal Killing vector field $K$ is used in the construction of the following symmetry operator of the twistor equation \cite{Ertem1}
\begin{equation}
{\cal{L}}_K-\frac{1}{2}\lambda
\end{equation}
where $\lambda$ is the function appearing in the definition of $K$ with respect to the metric $g$; ${\cal{L}}_Kg=2\lambda g$. Moreover, CKY forms which are antisymmetric generalizations of conformal Killing vector fields to higher-degree forms can also be used in the construction of more general symmetry operators of the twistor equation in constant curvature and Einstein manifolds. A $p$-form $\omega$ is called a CKY $p$-form, if it satisfies the following CKY equation in $n$-dimensions
\begin{equation}
\nabla_X\omega=\frac{1}{p+1}i_Xd\omega-\frac{1}{n-p+1}\widetilde{X}\wedge\delta\omega.
\end{equation}
The integrability condition of the CKY equation is given by
\begin{equation}
\frac{p}{p+1}\delta d\omega+\frac{n-p}{n-p+1}d\delta\omega=e^b\wedge i_{X^a}R(X_a, X_b)\omega.
\end{equation}
So, the following operator defined in terms of CKY $p$-forms is a symmetry operator of the twistor equation in constant curvature manifolds \cite{Ertem2}
\begin{equation}
L_{\omega}=-(-1)^p\frac{p}{n}\omega.\displaystyle{\not}D+\frac{p}{2(p+1)}d\omega+\frac{p}{2(n-p+1)}\delta\omega.
\end{equation}
Namely, if $\psi$ is a twistor spinor, then $L_{\omega}\psi$ also satisfies the twistor equation
\begin{equation}
\nabla_{X_a}L_{\omega}\psi=\frac{1}{n}e_a.\displaystyle{\not}DL_{\omega}\psi.
\end{equation}
On the other hand, the same operator defined in (49) is again a symmetry operator of the twistor equation in Einstein manifolds. However, in that case it is constructed out of normal CKY forms which are CKY forms that satisfy (47) and have the following integrability condition
\begin{equation}
\frac{p}{p+1}\delta d\omega+\frac{n-p}{n-p+1}d\delta\omega=-2(n-p)K_a\wedge i_{X^a}\omega
\end{equation}
where $K_a$ is defined in (44). Eq. (51) is a special case of (48) and all CKY forms are normal CKY forms in constant curvature manifolds. Indeed, the symmetry operator defined in (49) gives way to define the extended conformal superalgebras constructed out of twistor spinors and CKY forms by considering the graded Lie algebra structure of CKY forms and the higher-degree Dirac currents of twistor spinors as is shown in \cite{Ertem1,Ertem2,Ertem4}.

\subsection{Transformation operators}

By using twistor spinors, the operators that transform the solutions of the massless field equations with different spin to each other can be constructed \cite{Acik Ertem2}. For example, a spin raising operator that takes a solution of the massless spin-0 field equation and gives a solution of the massless spin-$\frac{1}{2}$ Dirac equation can be written in terms of twistor spinors \cite{Acik Ertem2, Penrose Rindler}. This operator can also be thought as an operator that transforms the twistor spinors to harmonic spinors by using the functions that satisfiy a massless field equation. Let us consider a function $f$ that satisfies the following conformal Laplace equation in $n$-dimensions
\begin{equation}
\Delta f-\frac{n-2}{4(n-1)}{\cal{R}}f=0
\end{equation}
and a twistor spinor $\psi$ that satisfies (40). Then, the following operator
\begin{equation}
L_f=\frac{n-2}{n}f\displaystyle{\not}D+df
\end{equation}
transforms the twistor spinors to harmonic spinors. Namely, if $\psi$ is a twistor spinor, then $L_f\psi$ satisfies the massless Dirac equation $\displaystyle{\not}DL_f\psi=0$. The proof of this statement can be seen as follows
\begin{eqnarray}
\displaystyle{\not}DL_f\psi&=&\displaystyle{\not}D\left(\frac{n-2}{n}f\displaystyle{\not}D\psi+df.\psi\right)\nonumber\\
&=&e^a.\nabla_{X_a}\left(\frac{n-2}{n}f\displaystyle{\not}D\psi+df.\psi\right)\nonumber\\
&=&e^a.\left(\frac{n-2}{n}(\nabla_{X_a}f)\displaystyle{\not}D\psi+\frac{n-2}{n}f\nabla_{X_a}\displaystyle{\not}D\psi+\nabla_{X_a}df.\psi+df.\nabla_{X_a}\psi\right)\nonumber\\
&=&\frac{n-2}{n}\left(\displaystyle{\not}df.\displaystyle{\not}D\psi+f\displaystyle{\not}D^2\psi\right)+\displaystyle{\not}ddf.\psi+e^a.df.\nabla_{X_a}\psi
\end{eqnarray}
where we have used the definitions of Hodge-de Rham and Dirac operators in (9) and (19). Since $f$ is a function, it satisfies $\delta f=0$ and $\displaystyle{\not}df=df$ with $\displaystyle{\not}ddf=-\delta df=\Delta f$. From the equations (40) and (41), we obtain
\begin{eqnarray}
\displaystyle{\not}DL_f\psi&=&\frac{n-2}{n}\left(df.\displaystyle{\not}D\psi-\frac{n}{4(n-1)}{\cal{R}}f\psi\right)+\Delta f.\psi+\frac{1}{n}e^a.df.e_a.\displaystyle{\not}D\psi\nonumber\\
&=&\left(\Delta f-\frac{n-2}{4(n-1)}{\cal{R}}f\right).\psi\nonumber\\
&=&0
\end{eqnarray}
where we have used eq. (6) in the second line and eq. (52) in the third line.

The natural question to ask at this point is that if there are more general differential forms that transform twistor spinors to harmonic spinors other than the functions satisfying (52). A proposal for an operator of this type is given in \cite{Benn Kress} in terms of potential forms. We consider a general first-order operator written in terms of a $p$-form $\alpha$ and acting on a twistor spinor $\psi$
\begin{equation}
L_{\alpha}\psi=\alpha.\displaystyle{\not}D\psi+\Omega.\psi
\end{equation}
where $\Omega$ is an inhomogeneous Clifford form. We need the condition that $L_{\alpha}\psi$ must satisfy the massless Dirac equation $\displaystyle{\not}DL_{\alpha}\psi=0$. So, we can write explicitly from (56) as
\begin{eqnarray}
\displaystyle{\not}DL_{\alpha}\psi&=&e^a.\nabla_{X_a}L_{\alpha}\psi\nonumber\\
&=&e^a.\nabla_{X_a}\alpha.\displaystyle{\not}D\psi+e^a.\alpha.\nabla_{X_a}\displaystyle{\not}D\psi+e^a.\nabla_{X_a}\Omega.\psi+e^a.\Omega.\nabla_{X_a}\psi.
\end{eqnarray}
From the equality $e^a.\nabla_{X_a}\alpha=\displaystyle{\not}d\alpha$ and the equations (40) and (42), we have
\begin{eqnarray}
\displaystyle{\not}DL_{\alpha}\psi&=&\displaystyle{\not}d\alpha.\displaystyle{\not}D\psi+\frac{n}{2}e^a.\alpha.K_a.\psi+\displaystyle{\not}d\Omega.\psi+\frac{1}{n}e^a.\Omega.e_a.\displaystyle{\not}D\psi\nonumber\\
&=&\left(\displaystyle{\not}d\alpha+\frac{(n-2\Pi)}{n}\eta\Omega\right).\displaystyle{\not}D\psi+\left(\displaystyle{\not}d\Omega+\frac{n}{2}e^a.\alpha.K_a\right).\psi.
\end{eqnarray}
Here, we have used the generalization of eq. (6) that is $e^a.\Omega.e_a=(n-2\Pi)\eta\Omega$ where $\eta$ is the main automorphism of the exterior algebra defined after eq. (3) and $\Pi\Omega=e^a\wedge i_{X_a}\Omega$ gives the degree of the form. Hence, if $L_{\alpha}\psi$ would be a harmonic spinor, then we have two conditions that have to be satisfied by $\alpha$ and $\Omega$
\begin{eqnarray}
\displaystyle{\not}d\alpha+\frac{n-2\Pi}{n}\eta\Omega&=&0\\
\displaystyle{\not}d\Omega+\frac{n}{2}e^a.\alpha.K_a&=&0.
\end{eqnarray}
As a special case, in even dimensions $n=2k$, we can choose $\Omega$ as a $k$-form (a middle form) and (59) gives $d\alpha=-\delta\alpha$ and this is only possible for $\alpha=0$ since $\alpha$ is a homogeneous $p$-form. One can see from (60) that $\Omega$ is a harmonic form in that case, namely it satisfies $d\Omega=\delta\Omega=0$. So, from the operator (56), one can draw the conclusion that in even dimensions, for a harmonic middle form $\Omega$ and a twistor spinor $\psi$, the spinor $\Omega.\psi$ is a harmonic spinor.

In general, if we choose $\alpha$ in (56) as a non-zero $p$-form, then $\Omega$ is a sum of ($p+1$) and ($p-1$)-forms as can be seen from (59). So, eq. (59) can be written in the following form
\begin{equation}
\Omega=\frac{(-1)^pn}{n-2(p+1)}d\alpha-\frac{(-1)^pn}{n-2(p-1)}\delta\alpha.
\end{equation}
By applying the Hodge-de Rham operator $\displaystyle{\not}d$ to the both sides of (61), we obtain
\begin{eqnarray}
\displaystyle{\not}d\Omega&=&\frac{(-1)^pn}{n-2(p+1)}\displaystyle{\not}dd\alpha-\frac{(-1)^pn}{n-2(p-1)}\displaystyle{\not}d\delta\alpha\nonumber\\
&=&-\frac{(-1)^pn}{n-2(p+1)}\delta d\alpha-\frac{(-1)^pn}{n-2(p-1)}d\delta\alpha.
\end{eqnarray}
By substituting (62) in (60), the conditions to obtain a harmonic spinor turn into the following equation
\begin{equation}
\frac{(-1)^p}{n-2(p+1)}\delta d\alpha+\frac{(-1)^p}{n-2(p-1)}d\delta\alpha=\frac{1}{2}e^a.\alpha.K_a.
\end{equation}
The right hand side of (63) can be written in a more explicit way as
\begin{eqnarray}
e^a.\alpha.K_a&=&(e^a\wedge\alpha+i_{X^a}\alpha).K_a\nonumber\\
&=&(-1)^p(i_{X^a}K_a)\alpha-2(-1)^pK_a\wedge i_{X^a}\alpha\nonumber\\
&=&(-1)^p\frac{2}{n-2}P_a\wedge i_{X^a}\alpha-(-1)^p\frac{n+2(p-1)}{2(n-1)(n-2)}{\cal{R}}\alpha
\end{eqnarray}
where we have used (44) in the last line. So, eq. (63) is written as follows
\begin{eqnarray}
\frac{1}{n-2(p+1)}\delta d\alpha+\frac{1}{n-2(p-1)}d\delta\alpha&=&\frac{1}{n-2}P_a\wedge i_{X^a}\alpha\nonumber\\
&&-\frac{n+2(p-1)}{4(n-1)(n-2)}{\cal{R}}\alpha.
\end{eqnarray}
The solutions of this equation are called potential forms. The name derives from the properties of the solutions for the special choices of $p$. For example, in even dimensions, for $p=\frac{n}{2}-1$, $\alpha$ is a potential form for the middle-form Maxwell equations and for $p=\frac{n}{2}+1$ it is a co-potential form for the same set of equations \cite{Benn Kress,Benn Charlton Kress}. Especially, for $p=0$, eq. (65) reduces to (52), so it is a generalization of (52) to higher-degree forms. Similar to eq. (52), (65) has also conformal covariance, namely it is covariant under transformations of the metric $g\rightarrow e^{2\lambda}g$ and $p$-forms $\alpha\rightarrow e^{(p+1-n/2)\lambda}\alpha$ for any function $\lambda$ \cite{Benn Kress}. So, (65) can also be called as the generalized conformal Laplace equation.

Then, we obtain the result that for a potential $p$-form $\alpha$ which satisfies (65) and a twistor spinor $\psi$, the spinor
\begin{equation}
L_{\alpha}\psi=\alpha.\displaystyle{\not}D\psi+\frac{(-1)^pn}{n-2(p+1)}d\alpha.\psi-\frac{(-1)^pn}{n-2(p-1)}\delta\alpha.\psi
\end{equation}
is a harmonic spinor.

\subsection{General picture}

CKY forms defined in (47) and used in the construction of the symmetry operators of twistor spinors in constant curvature and Einstein manifolds appear also in the construction of symmetry operators of the massless Dirac equation in all backgrounds \cite{Benn Charlton}. So, one can obtain a harmonic spinor from a known harmonic spinor $\psi$ and a CKY $p$-form $\omega$ by using the following symmetry operator
\begin{equation}
{\cal{L}}_{\omega}\psi=e^a.\omega.\nabla_{X_a}\psi+\frac{p}{p+1}d\omega.\psi-\frac{n-p}{n-p+1}\delta\omega.\psi.
\end{equation}
Namely, we have $\displaystyle{\not}D{\cal{L}}_{\omega}\psi=0$. A similar construction of symmetry operators for the massive Dirac equation can be done by using a subset of CKY forms that are called Killing-Yano (KY) forms which are antisymmetric generalizations of Killing vector fields to higher degree forms and satisfy the condition $\delta\omega=0$ \cite{Benn Kress2,Acik Ertem Onder Vercin2}.

In that way, from the investigations performed up to this point, we obtain three sets of transformation operators that take twistor spinors and give harmonic spinors. The first one is the first-order transformation operator from twistor spinors to harmonic spinors via potential forms $\alpha$ defined in (66)
\begin{equation}
\psi\longrightarrow L_{\alpha}\psi.
\end{equation}
The second set of operators are the second-order transformation operators which are the combinations of symmetry operators defined for twistors spinors in (49) and for harmonic spinors in (67) with the transformation operator in (66)
\begin{equation}
\psi\longrightarrow L_{\alpha}L_{\omega}\psi\qquad,\qquad\psi\longrightarrow{\cal{L}}_{\omega}L_{\alpha}\psi.
\end{equation}
The third set of operators are general third-order transformation operators from twistor spinors to harmonic spinors via CKY forms $\omega$ and potential forms $\alpha$
\begin{equation}
\psi\longrightarrow {\cal{L}}_{\omega}L_{\alpha}L_{\omega}\psi.
\end{equation}
Indeed, these second-order and third-order operators all reduce to other first-order operators since the integrability conditions (41-42) of twistor spinors show that the second-order derivatives of twistor spinors can be written in terms of the curvature characteristics of the manifold.

The general picture of the transformation properties of twistor spinors and harmonic spinors can be summarized as follows

\[
\textrm{twistor spinors}\quad\autorightarrow{$L_{\omega}$}{CKY forms}\quad\textrm{twistor spinors}
\]

\[
\textrm{twistor spinors}\quad\autorightarrow{$L_{\alpha}$}{potential forms}\quad\textrm{harmonic spinors}
\]

\[
\textrm{harmonic spinors}\quad\autorightarrow{${\cal{L}}_{\omega}$}{CKY forms}\quad\textrm{harmonic spinors}
\]

\section{Gauged twistors to gauged harmonic spinors}

In this section, we generalize the analysis in the previous section to $Spin^c$ spinors and gauged connections. The supersymmetry generators of supersymmetric field theories coupled with supergravity are given by $Spin^c$ spinors called gauged twistor spinors \cite{de Medeiros,Cassani Klare Martelli Tomasiello Zaffaroni}. Gauged twistor spinors are solutions of the following gauged twistor equation written in terms of the gauged connection $\widehat{\nabla}$ and the gauged Dirac operator $\widehat{\displaystyle{\not}D}$ defined in (24) and (35)
\begin{equation}
\widehat{\nabla}_X\psi=\frac{1}{n}\widetilde{X}.\widehat{\displaystyle{\not}D}\psi.
\end{equation}
Similar to the twistor spinors case, by applying the gauged covariant derivative operator to eq. (71), the integrability conditions of the gauged twistor equation can be obtained. The resulting integrability conditions are given as follows \cite{Ertem3}
\begin{eqnarray}
\widehat{\displaystyle{\not}D}^2\psi&=&-\frac{n}{4(n-1)}{\cal{R}}\psi+\frac{n}{n-1}F.\psi\\
\widehat{\nabla}_{X_a}\widehat{\displaystyle{\not}D}\psi&=&\frac{n}{2}K_a.\psi-\frac{n}{(n-1)(n-2)}e_a.F.\psi+\frac{n}{n-2}i_{X_a}F.\psi\\
C_{ab}.\psi&=&2(i_{X_a}i_{X_b}F)\psi+\frac{n}{n-2}\left(e_b.i_{X_a}F-e_a.i_{X_b}F\right).\psi\nonumber\\
&&+\frac{4}{(n-1)(n-2)}e_a.e_b.F.\psi
\end{eqnarray}
where $F=dA$ is the gauge curvature 2-form. For the ungauged case $A=0$, these reduce to the integrability conditions of twistor spinors in (41-43). The gauge curvature $F$ can be determined by the last integrability condition (74) written in terms of the conformal 2-forms $C_{ab}$ \cite{Cassani Martelli,de Medeiros2}. For example, in constant curvature manifolds, the conformal 2-forms vanishes $C_{ab}=0$ and this implies that the gauge curvature must be zero $F=0$. So, this gives the result that the gauged twistor spinors defined with respect to flat connections ($A\neq 0$ and $F=0$) can exist in constant curvature manifolds.

Symmetry operators of gauged twistor spinors can also be constructed from CKY forms in constant curvature manifolds in a similar way to the case of twistor spinors. The following operator written in terms of the gauged Dirac operator $\widehat{\displaystyle{\not}D}$ and CKY $p$-forms $\omega$
\begin{equation}
\widehat{L}_{\omega}=-(-1)^p\frac{p}{n}\omega.\widehat{\displaystyle{\not}D}+\frac{p}{2(p+1)}d\omega+\frac{p}{2(n-p+1)}\delta\omega
\end{equation}
is a symmetry operator for the gauged twistor spinors which is proved in \cite{Ertem3}. This is only true for constant curvature manifolds and in that case we have flat connections with $F=0$. So, for a gauged twistor spinor $\psi$ and a CKY $p$-form $\omega$ in a constant curvature manifold, $\widehat{L}_{\omega}\psi$ satisfies the gauged twistor equation
\begin{equation}
\widehat{\nabla}_ {X_a}\widehat{L}_{\omega}\psi=\frac{1}{n}e_a.\widehat{\displaystyle{\not}D}\widehat{L}_{\omega}\psi.
\end{equation}
Note that, we do not need to use gauged CKY forms, which are defined as the solutions of the gauged CKY equation constructed from the gauged connection, in the construction of symmetry operators and using CKY forms is enough to construct them. On the other hand, since the spinor bilinears of twistor spinors correspond to CKY forms, the symmetry operators defined in (75) can also be used for generating gauged twistor spinors from ordinary twistor spinors \cite{Ertem3}.

\subsection{Transformation operators}

One can also search for the transformation operators that take gauged twistor spinors and give gauged harmonic spinors as in the case of ungauged spinors. We start with a gauged twistor spinor $\psi$ that satisfies (71) and we also demand that it is an eigenspinor of the gauge curvature 2-form $F$, namely $F.\psi=\gamma\psi$ where $\gamma$ is a real number. Then, we consider a function $f$ that is a solution of the following gauged Laplace equation written in $n$-dimensions
\begin{equation}
\widehat{\Delta}f+\left[\left(1+\frac{n-2}{n-1}\right)\gamma-\frac{n-2}{4(n-1)}{\cal{R}}\right]f=0
\end{equation}
where $\gamma$ is the real number defined above. We propose that the following operator written in terms of the gauged Dirac operator $\widehat{\displaystyle{\not}D}$ and the gauged exterior derivative operator $\widehat{d}$ as
\begin{equation}
\widehat{L}_f=\frac{n-2}{n}f\widehat{\displaystyle{\not}D}+\widehat{d}f
\end{equation}
is an operator that transforms gauged twistor spinors to gauged harmonic spinors. So, for a gauged twistor spinor $\psi$, which is also an eigenspinor of $F$, $L_f\psi$ satisfies the gauged harmonic spinor equation $\widehat{\displaystyle{\not}D}\widehat{L}_f\psi=0$. We can prove this as follows
\begin{eqnarray}
\widehat{\displaystyle{\not}D}\widehat{L}_f\psi&=&e^a.\widehat{\nabla}_{X_a}\left(\frac{n-2}{n}f\widehat{\displaystyle{\not}D}\psi+\widehat{d}f.\psi\right)\nonumber\\
&=&e^a.\left(\frac{n-2}{n}(\widehat{\nabla}_{X_a}f)\widehat{\displaystyle{\not}D}\psi+\frac{n-2}{n}f\widehat{\nabla}_{X_a}\widehat{\displaystyle{\not}D}\psi+\widehat{\nabla}_{X_a}\widehat{d}f.\psi+\widehat{d}f.\widehat{\nabla}_{X_a}\psi\right)\nonumber\\
&=&\frac{n-2}{n}\left(\widehat{\displaystyle{\not}d}f.\widehat{\displaystyle{\not}D}\psi+f\widehat{\displaystyle{\not}D}^2\psi\right)+\widehat{\displaystyle{\not}d}\widehat{d}f.\psi+e^a.\widehat{\displaystyle{\not}d}f.\widehat{\nabla}_{X_a}\psi
\end{eqnarray}
where we have used the definitions of gauged Hodge-de Rham and gauged Dirac operators given in (29) and (35). Since $f$ is a function, we have $\widehat{\displaystyle{\not}d}f=\widehat{d}f$ and $\widehat{\displaystyle{\not}d}\widehat{d}f=\widehat{d}^2f-\widehat{\delta}\widehat{d}f=\widehat{d}^2f+\widehat{\Delta}f$. So, (79) can be written as
\begin{eqnarray}
\widehat{\displaystyle{\not}D}\widehat{L}_f\psi&=&\frac{n-2}{n}\left(\widehat{d}f.\widehat{\displaystyle{\not}D}\psi+f\widehat{\displaystyle{\not}D}^2\psi\right)+\widehat{d}^2f.\psi+\widehat{\Delta}f.\psi+e^a.\widehat{d}f.\widehat{\nabla}_{X_a}\psi\\
&=&\frac{n-2}{n}\left(\widehat{d}f.\widehat{\displaystyle{\not}D}\psi+f\widehat{\displaystyle{\not}D}^2\psi\right)+fF.\psi+\widehat{\Delta}f.\psi-\widehat{d}f.\widehat{\displaystyle{\not}D}\psi+2(i_{X^a}\widehat{d}f)\widehat{\nabla}_{X_a}\psi\nonumber
\end{eqnarray}
where we have used eq. (30) and $e^a.\widehat{d}f+\widehat{d}f.e^a=2i_{X^a}\widehat{d}f$ in the last line. Since $(i_{X^a}\widehat{d}f)\widehat{\nabla}_{X_a}=\widehat{\nabla}_{\widetilde{\widehat{d}f}}$ and $\psi$ satisfies $F.\psi=\gamma\psi$, we obtain
\begin{equation}
\widehat{\displaystyle{\not}D}\widehat{L}_f\psi=\widehat{\Delta}f.\psi+\gamma f\psi+\frac{n-2}{n}f\widehat{\displaystyle{\not}D}^2\psi+2\left(\widehat{\nabla}_{\widetilde{\widehat{d}f}}\psi-\frac{1}{n}\widehat{d}f.\widehat{\displaystyle{\not}D}\psi\right)
\end{equation}
and from the gauged twistor equation (71) and the integrability condition in (72) we reach the result
\begin{eqnarray}
\widehat{\displaystyle{\not}D}\widehat{L}_f\psi&=&\widehat{\Delta}f.\psi+\left[\left(1+\frac{n-2}{n-1}\right)\gamma-\frac{n-2}{4(n-1)}{\cal{R}}\right]f\psi\nonumber\\
&=&0
\end{eqnarray}
where we have used eq. (77) in the last line.

As we have discussed above, in constant curvature manifolds, the gauge curvature 2-form $F$ vanishes. So, the condition on the gauged twistor spinor $\psi$ to be an eigenspinor of $F$ is not necessary in that case. So, we can transform all gauged twistor spinors in constant curvature manifolds to gauged harmonic spinors by using the operator $L_f$ defined in (78) where $f$ satisfies the following gauged conformal Laplace equation
\begin{equation}
\widehat{\Delta}f-\frac{n-2}{4(n-1)}{\cal{R}}f=0
\end{equation}
which reduces from (77) by taking $\gamma=0$ and is a direct generalization of (52) to gauged connections.

We can search for transformation operators that take gauged twistor spinors and give gauged harmonic spinors and generalize the operator (78) to general differential forms. To obtain such an operator, we consider the following first-order operator acting on a gauged twistor spinor $\psi$ and is written in terms of the gauged Dirac operator $\widehat{\displaystyle{\not}D}$ and an arbitrary $p$-form $\alpha$
\begin{equation}
\widehat{L}_{\alpha}\psi=\alpha.\widehat{\displaystyle{\not}D}\psi+\Omega.\psi
\end{equation}
where $\Omega$ is an arbitrary inhomogeneous Clifford form. We need to find the conditions on $\alpha$ and $\Omega$ to have a gauged harmonic spinor, namely to satisfy the equation $\widehat{\displaystyle{\not}D}\widehat{L}_{\alpha}\psi=0$. By applying $\widehat{\displaystyle{\not}D}$ to (84) we obtain
\begin{eqnarray}
\widehat{\displaystyle{\not}D}\widehat{L}_{\alpha}\psi&=&e^a.\widehat{\nabla}_{X_a}\left(\alpha.\widehat{\displaystyle{\not}D}\psi+\Omega.\psi\right)\nonumber\\
&=&e^a.\left(\widehat{\nabla}_{X_a}\alpha.\widehat{\displaystyle{\not}D}\psi+\alpha.\widehat{\nabla}_{X_a}\widehat{\displaystyle{\not}D}\psi+\widehat{\nabla}_{X_a}\Omega.\psi+\Omega.\widehat{\nabla}_{X_a}\psi\right)\nonumber\\
&=&\widehat{\displaystyle{\not}d}\alpha.\widehat{\displaystyle{\not}D}\psi+e^a.\alpha.\widehat{\nabla}_{X_a}\widehat{\displaystyle{\not}D}\psi+\widehat{\displaystyle{\not}d}\Omega.\psi+e^a.\Omega.\widehat{\nabla}_{X_a}\psi
\end{eqnarray}
where we have used (29) in the last line. From the gauged twistor equation (71) and the integrability condition (73), we can write
\begin{eqnarray}
\widehat{\displaystyle{\not}D}\widehat{L}_{\alpha}\psi&=&\widehat{\displaystyle{\not}d}\alpha.\widehat{\displaystyle{\not}D}\psi+\frac{n}{2}e^a.\alpha.K_a.\psi-\frac{n}{(n-1)(n-2)}e^a.\alpha.e_a.F.\psi\nonumber\\
&&+\frac{n}{n-2}e^a.\alpha.i_{X_a}F.\psi+\widehat{\displaystyle{\not}d}\Omega.\psi+\frac{1}{n}e^a.\Omega.e_a.\widehat{\displaystyle{\not}D}\psi\nonumber\\
&=&\left(\widehat{\displaystyle{\not}d}\alpha+\frac{n-2\Pi}{n}\eta\Omega\right).\widehat{\displaystyle{\not}D}\psi\\
&&+\left(\widehat{\displaystyle{\not}d}\Omega+\frac{n}{2}e^a.\alpha.K_a-\frac{(-1)^p(n-2p)}{(n-1)(n-2)}\alpha.F+\frac{n}{n-2}e^a.\alpha.i_{X_a}F\right).\psi\nonumber
\end{eqnarray}
where we have used eq. (6) in the last line. So, to obtain a gauged harmonic spinor we have the following two conditions to be satisfied
\begin{eqnarray}
\widehat{\displaystyle{\not}d}\alpha+\frac{n-2\Pi}{n}\eta\Omega&=&0\\
\widehat{\displaystyle{\not}d}\Omega+\frac{n}{2}e^a.\alpha.K_a-\frac{(-1)^p(n-2p)}{(n-1)(n-2)}\alpha.F+\frac{n}{n-2}e^a.\alpha.i_{X_a}F&=&0.
\end{eqnarray}
For the special case of even dimensions, $n=2k$, we can choose $\Omega$ a $k$-form (a middle form) and obtain $\alpha=0$ from (87). Then, (88) gives that $\Omega$ must be a gauged harmonic form that satisfies the equalities $\widehat{d}\Omega=0$ and $\widehat{\delta}\Omega=0$. So, from the operator defined in (84), we conclude the fact that for a gauged twistor spinor $\psi$ and a gauged harmonic middle form $\Omega$, the spinor $\Omega.\psi$ corresponds to a gauged harmonic spinor.

For the general case of arbitrary dimensions and $\alpha$ a non-zero $p$-form, eq. (87) gives $\Omega$ as a sum of $(p+1)$ and $(p-1)$-forms as follows
\begin{equation}
\Omega=\frac{(-1)^pn}{n-2(p+1)}\widehat{d}\alpha-\frac{(-1)^pn}{n-2(p-1)}\widehat{\delta}\alpha.
\end{equation}
By applying $\widehat{\displaystyle{\not}d}$ to both sides of (89) we obtain
\begin{eqnarray}
\widehat{\displaystyle{\not}d}\Omega&=&\frac{(-1)^pn}{n-2(p+1)}\widehat{\displaystyle{\not}d}\widehat{d}\alpha-\frac{(-1)^pn}{n-2(p-1)}\widehat{\displaystyle{\not}d}\widehat{\delta}\alpha\nonumber\\
&=&\frac{(-1)^pn}{n-2(p+1)}\left(\widehat{d}^2\alpha-\widehat{\delta}\widehat{d}\alpha\right)-\frac{(-1)^pn}{n-2(p-1)}\left(\widehat{d}\widehat{\delta}\alpha-\widehat{\delta}^2\alpha\right)
\end{eqnarray}
where we have used $\widehat{\displaystyle{\not}d}=\widehat{d}-\widehat{\delta}$ in the last line. So, by substituting (88) in (90), one can write
\begin{eqnarray}
&&\frac{(-1)^pn}{n-2(p+1)}\left(\widehat{d}^2\alpha-\widehat{\delta}\widehat{d}\alpha\right)-\frac{(-1)^pn}{n-2(p-1)}\left(\widehat{d}\widehat{\delta}\alpha-\widehat{\delta}^2\alpha\right)\nonumber\\
&&=-\frac{n}{2}e^a.\alpha.K_a+\frac{(-1)^p(n-2p)}{(n-1)(n-2)}\alpha.F+\frac{n}{n-2}e^a.\alpha.i_{X_a}F.
\end{eqnarray}
By using the equalities (30) and (31) and the definition of $K_a$ with the calculation given in (64), we reach to the condition for obtaining a gauged harmonic spinor as follows
\begin{eqnarray}
\frac{1}{n-2(p+1)}\widehat{\delta}\widehat{d}\alpha+\frac{1}{n-2(p-1)}\widehat{d}\widehat{\delta}\alpha&=&\frac{1}{n-2}P_a\wedge i_{X^a}\alpha-\frac{n-2(p-1)}{4(n-1)(n-2)}{\cal{R}}\alpha\nonumber\\
&-&\frac{n-2p}{n(n-1)(n-2)}\alpha.F-\frac{(-1)^p}{n-2}e^a.\alpha.i_{X_a}F\nonumber\\
&-&\frac{1}{n-2(p-1)}(i_{X^a}i_{X^b}F)i_{X_a}i_{X_b}\alpha\nonumber\\
&+&\frac{1}{n-2(p+1)}F\wedge\alpha.
\end{eqnarray}
The terms in the second line which are written as Clifford products can be expanded in terms of wedge products by using the conventions defined in section II. Hence, we have the following expressions
\begin{equation}
\alpha.F=F\wedge\alpha+i_{X_a}F\wedge i_{X^a}\alpha-\frac{1}{2}(i_{X_a}i_{X_b}F)i_{X^a}i_{X^b}\alpha,
\end{equation}
\begin{eqnarray}
e^a.\alpha.i_{X_a}F&=&2(-1)^p\alpha\wedge F-2(-1)^pi_{X_a}F\wedge i_{X^a}\alpha\nonumber\\
&&+(-1)^p(i_{X_a}i_{X_b}F)i_{X^a}i_{X^b}\alpha.
\end{eqnarray}
The condition (92) transforms into the following form after substituting (93) and (94) in it
\begin{eqnarray}
&&\frac{1}{n-2(p+1)}\widehat{\delta}\widehat{d}\alpha+\frac{1}{n-2(p-1)}\widehat{d}\widehat{\delta}\alpha\nonumber\\
&=&\frac{1}{n-2}P_a\wedge i_{X^a}\alpha-\frac{n-2(p-1)}{4(n-1)(n-2)}{\cal{R}}\alpha\nonumber\\
&+&\left(\frac{2}{n-2}-\frac{n-2p}{n(n-1)(n-2)}\right)i_{X_a}F\wedge i_{X^a}\alpha\nonumber\\
&+&\left(\frac{1}{n-2(p+1)}-\frac{n-2p}{n(n-1)(n-2)}-\frac{2}{n-2}\right)F\wedge\alpha\nonumber\\
&-&\left(\frac{1}{n-2(p-1)}-\frac{n-2p}{2n(n-1)(n-2)}+\frac{1}{n-2}\right)(i_{X_a}i_{X_b}F)i_{X^a}i_{X^b}\alpha.
\end{eqnarray}
The left hand side of (95) is a $p$-form and the first two lines of the right hand side are again $p$-forms. However, the last two lines of the right hand side are $(p+2)$-form and $(p-2)$-form, respectively. So, for the consistency of the condition (95), the last two lines of the right hand side must be zero, and this is possible for $F=0$. Since the gauged connections are constructed out of the flat gauge potential 1-forms $A$ in constant curvature manifolds as is discussed at the beginning of this section, the condition (95) can be satisfied consistently in constant curvature manifolds. In that case, the condition (95) to obtain a gauged harmonic spinor can be written as follows
\begin{eqnarray}
\frac{1}{n-2(p+1)}\widehat{\delta}\widehat{d}\alpha+\frac{1}{n-2(p-1)}\widehat{d}\widehat{\delta}\alpha&=&\frac{1}{n-2}P_a\wedge i_{X^a}\alpha-\frac{n+2(p-1)}{4(n-1)(n-2)}{\cal{R}}\alpha\nonumber\\
&=&\left(\frac{p}{n(n-2)}-\frac{n+2(p-1)}{4(n-1)(n-2)}\right){\cal{R}}\alpha\nonumber\\
\end{eqnarray}
where we have used the relations $P_a=\frac{\cal{R}}{n}e_a$ in constant curvature manifolds and $e^a\wedge i_{X_a}\alpha=p\alpha$ in the last line. The solutions of eq. (96) are called gauged potential forms and it is a direct generalization of eq. (65) to the gauged exterior derivative $\widehat{d}$ and the gauged co-derivative $\widehat{\delta}$. Because of the conformal covariance of the first line of (96), the construction might also be extended to conformally-flat manifolds.

So, we prove the fact that a gauged twistor spinor $\psi$ can be transformed into a gauged harmonic spinor by using a gauged potential form $\alpha$ through the following transformation operator
\begin{equation}
\widehat{L}_{\alpha}\psi=\alpha.\widehat{\displaystyle{\not}D}\psi+\frac{(-1)^pn}{n-2(p+1)}\widehat{d}\alpha.\psi-\frac{(-1)^pn}{n-2(p-1)}\widehat{\delta}\alpha.\psi
\end{equation}
in constant curvature manifolds.

\subsection{Symmetry operators of gauged harmonic spinors}

One can also search for the symmetry operators of the massless gauged Dirac equation that transform gauged harmonic spinors to each other. In the light of the ungauged case, we propose the following operator written in terms of the gauged quantities as a symmetry operator of gauged harmonic spinors
\begin{equation}
\widehat{\cal{L}}_{\omega}=e^a.\omega.\widehat{\nabla}_{X_a}+\frac{p}{p+1}\widehat{d}\omega-\frac{n-p}{n-p+1}\widehat{\delta}\omega
\end{equation}
where $\omega$ is a $p$-form which satisfies the following gauged CKY equation
\begin{equation}
\widehat{\nabla}_{X}\omega=\frac{1}{p+1}i_X\widehat{d}\omega-\frac{1}{n-p+1}\widetilde{X}\wedge\widehat{\delta}\omega
\end{equation}
for any vector field $X$. The integrability condition of the gauged CKY equation is given by
\begin{equation}
\frac{p}{p+1}\widehat{\delta}\widehat{d}\omega+\frac{n-p}{n-p+1}\widehat{d}\widehat{\delta}\omega=e^b\wedge i_{X^a}\widehat{R}(X_a, X_b)\omega.
\end{equation}
So, for a gauged harmonic spinor $\psi$, we have to check the condition that $\widehat{\cal{L}}_{\omega}\psi$ is again a gauged harmonic spinor, namely we have $\widehat{\displaystyle{\not}D}\widehat{\cal{L}}_{\omega}\psi=0$. By applying $\widehat{\cal{L}}_{\omega}$ and $\widehat{\displaystyle{\not}D}$ to a gauged harmonic spinor $\psi$, we obtain
\begin{eqnarray}
\widehat{\displaystyle{\not}D}\widehat{\cal{L}}_{\omega}\psi&=&e^b.\widehat{\nabla}_{X_b}\left(e^a.\omega.\widehat{\nabla}_{X_a}\psi+\frac{p}{p+1}\widehat{d}\omega.\psi-\frac{n-p}{n-p+1}\widehat{\delta}\omega.\psi\right)\nonumber\\
&=&e^b.e^a.\widehat{\nabla}_{X_b}\omega.\widehat{\nabla}_{X_a}\psi+e^b.e^a.\omega.\widehat{\nabla}_{X_b}\widehat{\nabla}_{X_a}\psi\nonumber\\
&&+\frac{p}{p+1}e^b.\widehat{\nabla}_{X_b}\left(\widehat{d}\omega.\psi\right)-\frac{n-p}{n-p+1}e^b.\widehat{\nabla}_{X_b}\left(\widehat{\delta}\omega.\psi\right)
\end{eqnarray}
where we have used (25) for the normal coordinates in the second line. We will calculate the terms on the right hand side of (101) separately in an explicit way.

The first term on the right hand side of (101) can be written in the following way by using (1) and the definition (29)
\begin{eqnarray}
e^b.e^a.\widehat{\nabla}_{X_b}\omega.\widehat{\nabla}_{X_a}\psi&=&2\widehat{\nabla}_{X^a}\omega.\widehat{\nabla}_{X_a}\psi-e^a.\widehat{\displaystyle{\not}d}\omega.\widehat{\nabla}_{X_a}\psi\nonumber\\
&=&\frac{2}{p+1}i_{X^a}\widehat{d}\omega.\widehat{\nabla}_{X_a}\psi-\frac{2}{n-p+1}(e^a\wedge\widehat{\delta}\omega).\widehat{\nabla}_{X_a}\psi\nonumber\\
&&-e^a.\widehat{d}\omega.\widehat{\nabla}_{X_a}\psi+e^a.\widehat{\delta}\omega.\widehat{\nabla}_{X_a}\psi
\end{eqnarray}
where we have used the gauged CKY equation (99) in the second line and $\widehat{\displaystyle{\not}d}=\widehat{d}-\widehat{\delta}$ in the third line. From the expressions of the Clifford product of a 1-form and a $p$-form given in (2) and (3), we can write
\begin{eqnarray}
i_{X^a}\widehat{d}\omega&=&\frac{1}{2}\left(e^a.\widehat{d}\omega+(-1)^p\widehat{d}\omega.e^a\right)\\
e^a\wedge\widehat{\delta}\omega&=&\frac{1}{2}\left(e^a.\widehat{\delta}\omega-(-1)^p\widehat{\delta}\omega.e^a\right).
\end{eqnarray}
By substituting them in (102), we have
\begin{eqnarray}
e^b.e^a.\widehat{\nabla}_{X_b}\omega.\widehat{\nabla}_{X_a}\psi&=&-\frac{p}{p+1}e^a.\widehat{d}\omega.\widehat{\nabla}_{X_a}\psi+\frac{n-p}{n-p+1}e^a.\widehat{\delta}\omega.\widehat{\nabla}_{X_a}\psi\nonumber\\
&&+\frac{(-1)^p}{p+1}\widehat{d}\omega.\widehat{\displaystyle{\not}D}\psi+\frac{(-1)^p}{n-p+1}\widehat{\delta}\omega.\widehat{\displaystyle{\not}D}\psi\nonumber\\
&=&-\frac{p}{p+1}e^a.\widehat{d}\omega.\widehat{\nabla}_{X_a}\psi+\frac{n-p}{n-p+1}e^a.\widehat{\delta}\omega.\widehat{\nabla}_{X_a}\psi
\end{eqnarray}
where we have used the massless gauged Dirac equation (38) in the last line.

To calculate the second term on the right hand side of (101), we consider the relation $e^b.e^a=e^b\wedge e^a+i_{X^b}e^a$, then we can write
\begin{eqnarray}
e^b.e^a.\omega.\widehat{\nabla}_{X_b}\widehat{\nabla}_{X_a}\psi&=&(e^b\wedge e^a).\omega.\widehat{\nabla}_{X_b}\widehat{\nabla}_{X_a}\psi+\omega.\widehat{\nabla}_{X^a}\widehat{\nabla}_{X_a}\psi\nonumber\\
&=&\frac{1}{2}e^{ba}.\omega.\widehat{R}(X_b, X_a)\psi+\omega.\widehat{\nabla}^2\psi
\end{eqnarray}
where $e^{ba}=e^b\wedge e^a$ and we consider the antisymmetry of $e^{ba}$ to take the antisymmetric part of the first term on the right hand side and use the definition of the gauged curvature operator (33) and the gauged version of the trace of the Hessian (16) in normal coordinates. By writing the gauged curvature operator in terms of the ungauged one as in (34) and using the equality (36), eq. (106) turns into
\begin{eqnarray}
e^b.e^a.\omega.\widehat{\nabla}_{X_b}\widehat{\nabla}_{X_a}\psi&=&\frac{1}{2}e^{ba}.\omega.R(X_b, X_a)\psi+F.\omega.\psi+\frac{1}{4}{\cal{R}}\omega.\psi-\omega.F.\psi\nonumber\\
&=&\frac{1}{4}e^{ab}.\omega.R_{ab}.\psi+\frac{1}{4}{\cal{R}}\omega.\psi+[F, \omega]_{Cl}.\psi\nonumber\\
&=&\frac{1}{4}e^{ab}.\omega.R_{ab}.\psi+\frac{1}{4}{\cal{R}}\omega.\psi-2\left(i_{X_a}F\wedge i_{X^a}\omega\right).\psi
\end{eqnarray}
where we have used $e^{ba}(i_{X_a}i_{X_b}F)=2F$ and $\widehat{\displaystyle{\not}D}\psi=0$ in the first line, eq. (20) in the second line and eq. (5) in the third line. Because of the pairwise symmetry of the components of the Riemann tensor, namely $R_{abcd}=R_{cdab}$, we have $e^{ab}.\omega.R_{ab}=R_{ab}.\omega.e^{ab}$. Moreover, this term can also be written in terms of the curvature endomorphism as given in (14) and we obtain
\begin{eqnarray}
e^b.e^a.\omega.\widehat{\nabla}_{X_b}\widehat{\nabla}_{X_a}\psi&=&e^a\wedge i_{X^b}R(X_b, X_a)\omega.\psi-2\left(i_{X_a}F\wedge i_{X^a}\omega\right).\psi\nonumber\\
&=&e^a\wedge i_{X^b}\widehat{R}(X_b, X_a)\omega.\psi-3\left(i_{X_a}F\wedge i_{X^a}\omega\right).\psi
\end{eqnarray}
where we have used (34) in the last line. So, from the integrability condition (100) of the gauged CKY equation, we can finally write
\begin{eqnarray}
e^b.e^a.\omega.\widehat{\nabla}_{X_b}\widehat{\nabla}_{X_a}\psi&=&\frac{p}{p+1}\widehat{\delta}\widehat{d}\omega.\psi+\frac{n-p}{n-p+1}\widehat{d}\widehat{\delta}\omega.\psi\nonumber\\
&&-3\left(i_{X_a}F\wedge i_{X^a}\omega\right).\psi.
\end{eqnarray}

The third term on the right hand side of (101) can be calculated explicitly as follows
\begin{eqnarray}
e^a.\widehat{\nabla}_{X_a}\left(\widehat{d}\omega.\psi\right)&=&e^a.\left(\widehat{\nabla}_{X_a}\widehat{d}\omega.\psi+\widehat{d}\omega.\widehat{\nabla}_{X_a}\psi\right)\nonumber\\
&=&\widehat{\displaystyle{\not}d}\widehat{d}\omega.\psi+e^a.\widehat{d}\omega.\widehat{\nabla}_{X_a}\psi\nonumber\\
&=&\widehat{d}^2\omega.\psi-\widehat{\delta}\widehat{d}\omega.\psi+e^a.\widehat{d}\omega.\widehat{\nabla}_{X_a}\psi\nonumber\\
&=&\left(F\wedge\omega\right).\psi-\widehat{\delta}\widehat{d}\omega.\psi+e^a.\widehat{d}\omega.\widehat{\nabla}_{X_a}\psi
\end{eqnarray}
where we have used $\widehat{\displaystyle{\not}d}=\widehat{d}-\widehat{\delta}$ and eq. (30). The calculation of the fourth term on the right hand side of (101) can also be maintained explicitly as
\begin{eqnarray}
e^a.\widehat{\nabla}_{X_a}\left(\widehat{\delta}\omega.\psi\right)&=&e^a.\left(\widehat{\nabla}_{X_a}\widehat{\delta}\omega.\psi+\widehat{\delta}\omega.\widehat{\nabla}_{X_a}\psi\right)\nonumber\\
&=&\widehat{\displaystyle{\not}d}\widehat{\delta}\omega.\psi+e^a.\widehat{\delta}\omega.\widehat{\nabla}_{X_a}\psi\nonumber\\
&=&\widehat{d}\widehat{\delta}\omega.\psi-\widehat{\delta}^2\omega.\psi+e^a.\widehat{\delta}\omega.\widehat{\nabla}_{X_a}\psi\nonumber\\
&=&\widehat{d}\widehat{\delta}\omega.\psi+(i_{X^a}i_{X^b}F)i_{X_a}i_{X_b}\omega.\psi+e^a.\widehat{\delta}\omega.\widehat{\nabla}_{X_a}\psi
\end{eqnarray}
where we have used eq. (31). Now, we can sum up (105), (109), (110) and (111) and write eq. (101) explicitly in the following form
\begin{eqnarray}
\widehat{\displaystyle{\not}D}\widehat{\cal{L}}_{\omega}\psi&=&\frac{p}{p+1}\left(F\wedge\omega\right).\psi-3\left(i_{X^a}F\wedge i_{X_a}\omega\right).\psi\nonumber\\
&&-\frac{n-p}{n-p+1}(i_{X^a}i_{X^b}F)i_{X_a}i_{X_b}\omega.\psi.
\end{eqnarray}
So, to obtain a gauged harmonic spinor, the right hand side of (112) must be zero and this implies $F=0$. This condition is achieved in constant curvature manifolds. Then, we prove the result that the operator $\widehat{\cal{L}}_{\omega}$ defined in (98) and written in terms of gauged CKY forms $\omega$ is a symmetry operator of gauged harmonic spinors with flat connection in constant curvature manifolds. In fact, for $F=0$, gauged CKY forms reduce to CKY forms since they can be written as a product of a function and a CKY form in that case. Hence, the symmetry operators of gauged harmonic spinors can be constructed from CKY forms $\omega$ in constant curvature manifolds. This is expected since the similar situation appears in the case of symmetry operators of gauged twistor spinors given in eq. (75) and \cite{Ertem3}.

\subsection{General picture}

In this section, we construct three sets of operators that transform gauged twistor spinors to gauged harmonic spinors which are relevant in constant curvature manifolds. These operators are generalizations of the operators defined in section III to the gauged case. The first set of operators are the first-order operators defined in (97) which are written in terms of gauged potential forms $\alpha$ and act to a gauged twistor spinor $\psi$
\begin{equation}
\psi\longrightarrow \widehat{L}_{\alpha}\psi.
\end{equation}
The second set of operators are the second-order operators which are the combinations of symmetry operators defined for gauged twistors spinors in (75) and for gauged harmonic spinors in (98) with the transformation operator in (97) and written in terms of gauged potential forms $\alpha$, CKY forms $\omega$ and $\omega '$
\begin{equation}
\psi\longrightarrow \widehat{L}_{\alpha}\widehat{L}_{\omega}\psi\qquad,\qquad\psi\longrightarrow\widehat{\cal{L}}_{\omega '}\widehat{L}_{\alpha}\psi.
\end{equation}
The third set of operators are general third-order transformation operators from gauged twistor spinors to gauged harmonic spinors via CKY forms $\omega$, $\omega '$ and gauged potential forms $\alpha$
\begin{equation}
\psi\longrightarrow \widehat{\cal{L}}_{\omega '}\widehat{L}_{\alpha}\widehat{L}_{\omega}\psi.
\end{equation}
As in the case of ungauged operators, these second-order and third-order operators also reduce to other first-order operators since the integrability conditions (72-73) of gauged twistor spinors show that the second-order derivatives of gauged twistor spinors can be written in terms of curvature characteristics of the manifold and the gauge curvature 2-form.

The general picture of the transformation operators of gauged twistor spinors and gauged harmonic spinors in constant curvature manifolds can be summarized as follows

\[
\textrm{gauged twistor spinors}\quad\autorightarrow{$\widehat{L}_{\omega}$}{CKY forms}\quad\textrm{gauged twistor spinors}
\]

\[
\textrm{gauged twistor spinors}\quad\autorightarrow{$\widehat{L}_{\alpha}$}{gauged potential forms}\quad\textrm{gauged harmonic spinors}
\]

\[
\textrm{gauged harmonic spinors}\quad\autorightarrow{$\widehat{\cal{L}}_{\omega '}$}{CKY forms}\quad\textrm{gauged harmonic spinors}
\]

\section{Seiberg-Witten solutions}

Gauged harmonic spinors play an important role in the solutions of the Seiberg-Witten equations. Seiberg-Witten equations determine the topological invariants that classify four-dimensional manifolds and they are written in terms of $Spin^c$ spinors with an abelian gauge connection \cite{Seiberg Witten}. Gauged harmonic spinor equation is one of the Seiberg-Witten equations which are given as follows in four dimensions
\begin{eqnarray}
\widehat{\displaystyle{\not}D}\psi&=&0\\
F^+&=&-\frac{1}{4}\tau^{\psi}
\end{eqnarray}
with respect to a $Spin^c$ spinor $\psi$ and an abelian gauge connection $A$. Here, $F^+$ denotes the self-dual part of the gauge curvature 2-form $F$. The 2-form $\tau^{\psi}$ is defined by \cite{Friedrich}
\begin{equation}
\tau^{\psi}=(e_a.e_b.\psi, \psi)e^a\wedge e^b
\end{equation}
where $(\,,\,)$ corresponds to the spinor inner product in four dimensions. Indeed, $\tau^{\psi}$ is written in terms of 2-form Dirac currents of $\psi$. $p$-form Dirac current of a spinor $\psi$ is defined as
\begin{equation}
(\psi\overline{\psi})_p=(e_{a_1}.e_{a_2}...e_{a_p}.\psi, \psi)e^{a_1}\wedge e^{a_2}\wedge ... \wedge e^{a_p}
\end{equation}
where $\overline{\psi}$ denotes the dual spinor. So, we have
\begin{equation}
\tau^{\psi}=(\psi\overline{\psi})_2.
\end{equation}

We obtain gauged harmonic spinors in section 4 by starting from twistor spinors and using CKY forms and potential forms. We can search for the algebraic conditions to obtain Seiberg-Witten solutions from the found gauged harmonic spinors. In constant curvature manifolds we have $F=0$ and the equation (117) turns into $\tau^{\psi}=0$ in that case. Then, by starting with a gauged twistor spinor $\psi$, we obtain solutions of the Seiberg-Witten equations $\widehat{L}_{\alpha}\psi$, $\widehat{L}_{\alpha}\widehat{L}_{\omega}\psi$, $\widehat{\cal{L}}_{\omega '}\widehat{L}_{\alpha}\psi$ or $\widehat{\cal{L}}_{\omega '}\widehat{L}_{\alpha}\widehat{L}_{\omega}\psi$, if they have vanishing 2-form Dirac currents, namely if they satisfy the following algebraic conditions
\begin{eqnarray}
(\widehat{L}_{\alpha}\psi\overline{\widehat{L}_{\alpha}\psi})_2&=&0\\
(\widehat{L}_{\alpha}\widehat{L}_{\omega}\psi\overline{\widehat{L}_{\alpha}\widehat{L}_{\omega}\psi})_2&=&0\\
(\widehat{\cal{L}}_{\omega '}\widehat{L}_{\alpha}\psi\overline{\widehat{\cal{L}}_{\omega '}\widehat{L}_{\alpha}\psi})_2&=&0\\
(\widehat{\cal{L}}_{\omega '}\widehat{L}_{\alpha}\widehat{L}_{\omega}\psi\overline{\widehat{\cal{L}}_{\omega '}\widehat{L}_{\alpha}\widehat{L}_{\omega}\psi})_2&=&0
\end{eqnarray}
where $\alpha$, $\omega$ and $\omega '$ are as in section IV.C. So, a gauged twistor spinor defined with respect to a flat connection can be transformed into a solution of the Seiberg-Witten equations, if it satisfies one of the conditions in (121-124);

\[
\textrm{gauged harmonic spinors}\quad\autorightarrow{$\tau^{\psi}=0$}{vanishing 2-form Dirac currents}\quad\textrm{Seiberg-Witten solutions}
\]

\section{Conclusion}

We explicitly show that twistor spinors and harmonic spinors can be transformed to each other by using the transformation operators constructed out of conformal Laplace functions and potential forms. Similar constructions are generalized to gauged twistor spinors and gauged harmonic spinors and we prove that the transformation operators are written in terms of generalized gauged Laplace functions and gauged potential forms in this case. We also find the symmetry operators of gauged harmonic spinors by using CKY forms which generate gauged harmonic spinors from the known ones. Solutions of the Seiberg-Witten equations can also be obtained by using the found transformation and symmetry operators that satisfy some extra algebraic conditions.

The explicit construction of transformation operators from twistors to harmonic spinors can be handled by finding the CKY forms and potential forms of the ambient manifold. Since the second-order and third-order transformation operators defined in the paper reduce to first-order ones because of the integrability conditions of the twistor and gauged twistor equations, the symmetry operators acting on potential forms and gauged potential forms can be investigated from them. Through the commuting set of symmetry operators, one can also search for the general solutions of the potential and gauged potential equations. In that way, the construction of transformation operators in various constant curvature manifolds can be studied.



\begin{references}

\bibitem{Baum Friedrich Grunewald Kath} H. Baum, T. Friedrich, R. Grunewald and I. Kath, \emph{Twistors and Killing Spinors on Riemannian Manifolds} (Teubner, Stuttgart/Leipzig, 1991).

\bibitem{Bourguignon et al} J. P. Bourguignon, O. Hijazi, J. L. Milhorat, A. Moroianu, S. Moroianu, \emph{A Spinorial Approach to Riemannian and Conformal Geometry} (European Mathematical Society, Zurich, 2015).

\bibitem{de Medeiros} P. de Medeiros, "Rigid supersymmetry, conformal coupling and twistor spinors," J. High Energy Phys. JHEP\textbf{1409}, 032 (2014).

\bibitem{Cassani Klare Martelli Tomasiello Zaffaroni} D. Cassani, C. Klare, D. Martelli, A. Tomasiello and A. Zaffaroni, "Supersymmetry in Lorentzian curved spaces and holography," Commun. Math. Phys. \textbf{327}, 577 (2014).

\bibitem{Acik Ertem1} \"{O}. A\c{c}{\i}k and \"{U}. Ertem, "Higher-degree Dirac currents of twistor and Killing spinors in supergravity theories," Class. Quantum Grav. \textbf{32}, 175007 (2015).

\bibitem{Benn Charlton} I. M. Benn and P. Charlton, "Dirac symmetry operators from conformal Killing-Yano tensors," Class. Quantum Grav. \textbf{14}, 1037 (1997).

\bibitem{Ertem1} \"{U}. Ertem, "Extended superalgebras from twistor and Killing spinors," Diff. Geom. Appl. \textbf{54}, 236 (2017).

\bibitem{Ertem2} \"{U}. Ertem, "Twistor spinors and extended conformal superalgebras," arXiv:1605.03361 [hep-th] (2016).

\bibitem{Benn Kress} I. M. Benn and J. Kress, "Differential forms relating twistors to Dirac fields," in: \emph{Differential Geometry and its Applications, Proceedings of the 10th International Conference DGA 2007} pp. 573. (World Scientific Publishing, Singapore, 2008).

\bibitem{Seiberg Witten} N. Seiberg and E. Witten, "Monopoles, duality and chiral symmetry breaking in N=2 supersymmetric QCD," Nucl. Phys. B \textbf{431}, 581 (1994).

\bibitem{Festuccia Seiberg} G. Festuccia and N. Seiberg, "Rigid supersymmetric theories in curved superspace". J. High Energy Phys. JHEP\textbf{1106}, 114 (2011).

\bibitem{Klare Tomasiello Zaffaroni} C. Klare, A. Tomasiello and A. Zaffaroni, "Supersymmetry on curved spaces and holography," J. High Energy Phys. JHEP\textbf{1208}, 061 (2012).

\bibitem{Cassani Martelli} D. Cassani and D. Martelli, "Supersymmetry on curved spaces and superconformal anomalies," J. High Energy Phys. JHEP\textbf{1310}, 025 (2013).

\bibitem{Ertem3} \"{U}. Ertem, "Gauged twistor spinors and symmetry operators," J. Math. Phys. \textbf{58}, 032302 (2017).

\bibitem{Benn Tucker} I. M. Benn and R. W. Tucker, \emph{An Introduction to Spinors and Geometry with Applications in Physics} (IOP Publishing, Bristol, 1987).

\bibitem{Charlton} P. Charlton, \emph{The Geometry of Pure Spinors, with Applications} PhD Thesis, University of Newcastle (1997).

\bibitem{Acik Ertem Onder Vercin} \"{O}. A\c{c}{\i}k, \"{U}. Ertem, M. \"{O}nder and A. Ver\c{c}in, "Basic gravitational currents and Killing-Yano forms," Gen. Relativ. Gravit. \textbf{42} 2543 (2010).

\bibitem{Ertem4} \"{U}. Ertem, "Lie algebra of conformal Killing-Yano forms," Class. Quantum Grav. \textbf{33}, 125033 (2016).

\bibitem{Acik Ertem2} \"{O}. A\c{c}{\i}k and \"{U}. Ertem, "Spin raising and lowering operators for Rarita-Schwinger fields," Phys. Rev. D \textbf{98}, 066004 (2018).

\bibitem{Penrose Rindler} R. Penrose and W. Rindler, \emph{Spinors and Space-time} vol. 2 (Cambridge University Press, Cambridge, 1986).

\bibitem{Benn Charlton Kress} I. M. Benn, P. Charlton and J. Kress, "Debye potentials for Maxwell and Dirac fields from a generalization of the Killing-Yano equation," J. Math. Phys. \textbf{38}, 4504 (1997).

\bibitem{Benn Kress2} I. M. Benn and J. Kress, "First-order Dirac symmetry operators," Class. Quantum Grav. \textbf{21}, 427 (2004).

\bibitem{Acik Ertem Onder Vercin2} \"{O}. A\c{c}{\i}k, \"{U}. Ertem, M. \"{O}nder and A. Ver\c{c}in, "First-order symmetries of the Dirac equation in a curved background: a unified dynamical symmetry condition," Class. Quantum Grav. \textbf{26}, 075001 (2009).

\bibitem{de Medeiros2} P. de Medeiros, "Submaximal conformal symmetry superalgebras for Lorentzian manifolds of low dimension," J. High Energy Phys. JHEP\textbf{1602}, 008 (2016).

\bibitem{Friedrich} T. Friedrich, \emph{Dirac Operators in Riemannian Geometry} (American Mathematical Society, 2000).

\end{references}
\end{document}